\documentclass[11pt,a4paper,reqno]{amsart}
\usepackage{multirow}
\usepackage[centertags]{amsmath}
\usepackage{amsfonts}
\usepackage{amssymb}
\usepackage{amsthm}
\usepackage{newlfont}
\usepackage{a4}
\usepackage[pdftex]{graphicx,color}

\theoremstyle{definition}

\theoremstyle{remark}

\usepackage{bbm}

\def\R{\mathbb R}
\def\Z{\mathbb Z}
\def\N{\mathbb N}
\def\C{\mathbb C}

\def\a{\alpha}
\def\o{\omega}
\newcommand{\ep}{\varepsilon}
\newcommand{\la}{\lambda}

\newcommand{\map}{\rightarrow}

\newcommand{\id}{\mathfrak{1}}

\newcommand{\sca}[2]{\langle #1,\, #2\rangle}
\newcommand{\set}[2]{\left\{#1 \, |\, #2 \right\}}

\begin{document}

\title{Four families of Weyl group orbit functions of $B_3$ and $C_3$.}

\author[L. H\'akov\'a]{Lenka H\'akov\'a$^{1}$}
\author[J. Hrivn\'{a}k]{Ji\v{r}\'{\i} Hrivn\'{a}k$^{1,2}$}
\author[J. Patera]{Ji\v{r}\'{\i} Patera$^{1,3}$}
\date{\today}

\begin{abstract}
\small 
The properties of the four families of special functions of
three real variables, called here $C-$, $S-$, $S^s-$ and $S^l-$functions, are studied. The $S^s-$ and $S^l-$functions are considered in all details required for their exploitation in Fourier expansions of digital data, sampled on finite fragment of lattices of any density and of the $3D$ symmetry imposed by the weight lattices of $B_3$ and $C_3$ simple Lie algebras/groups. The continuous interpolations, which are induced by the discrete
expansions, are exemplified and compared for some model functions.
\end{abstract}
\maketitle

{\small
\noindent
$^1$ Centre de Recherches Math\'ematiques et D\'epartement de Math\'ematiques et de Statistique,
         Universit\'e de Montr\'eal, C.~P.~6128 -- Centre Ville,
         Montr\'eal, H3C\,3J7, Qu\'ebec, Canada;\\ 
$^2$ Department of Physics, Faculty of Nuclear Sciences and
Physical Engineering, Czech Technical University in Prague,
B\v{r}ehov\'a~7, 115 19 Prague 1, Czech Republic;\\
$^3$ MIND Research Institute,
Irvine, CA 92617, USA

\medskip
\noindent
\textit{E-mail:} hakova@dms.umontreal.ca, jiri.hrivnak@fjfi.cvut.cz, patera@crm.umontreal.ca
}
%%%%%%%%%%%%%%%%%%%%%%%%%%%%%%%%%%%%%%%%%%%%%%%%%%%%%%%%%%%
\maketitle

%%%%%%%%%%%%%%%%%%%%%%%%%%%%%%%%%%%%%%%%%%%%%%%%%%%%%%%%%%%

\section{Introduction}

Four families of functions, depending on three real variables, called here as $C-$, $S-$, $S^s-$ and $S^l-$functions, are described. Each family is complete within its functional space \cite{MMP} and orthogonal when integrated over a finite region $F$ of real Euclidean space $\R^3$. Moreover, the functions of each family are discretely orthogonal on a fraction of a lattice in $F$ of density of our choice and of symmetry dictated by the simple Lie algebras $B_3$ and $C_3$.

The definition of these functions for any number of variables and some of their properties are described in~\cite{KP2,KP3, MMP}. The first two families of functions are also called $C-$ and $S-$functions. They are well known as constituents of irreducible characters of compact simple Lie groups. Indeed, the irreducible characters can be expressed as finite sums of $C-$functions with integer coefficients called dominant weight multiplicities. Ratio of two $S-$functions appears in the Weyl character formula \cite{Hum}. Two other families are called $S^l-$ and $S^s-$ functions. Inspired by the Weyl character formula, we can consider so called hybrid characters, i.e. the ratio of two $S^l-$ or $S^s-$ functions. It is possible to show that they decompose into the sums of $C-$functions with integer coefficients. The explicit formulas for coefficients are in~\cite{LPS}. 

It is well known that the $C-$functions play an important role in the definition of Jacobi polynomials \cite{HO,KP4,KP5}. One can also remark that the characters and the hybrid characters could be derived as special cases of Jacobi polynomials. However, important insight into the properties of these functions would have been lost in the generality of the approach. For example, their discrete orthogonality appears to be outside of that approach, which is a handicap in the world of ever growing amount of digital data. From our perspective particularly useful are the four families discretized within $F$. The problem of discretization of the $C-$ and $S-$functions, which is now over 20 years old~\cite{HP,MP1,MP84,MP2}, is carried over to the dicretization of the other two families in~\cite{HMP}.
	
Our motivation for studying these families of functions is guided by ever increasing need to process three-dimensional digital data. Discrete orthogonality of these functions open new efficient possibilities for precisely that. Moreover, the $3D$ symmetry imposed by the weight lattices of $B_3$ and $C_3$ should be advantageous in describing quantum systems possessing such a symmetry, as well as in some problems of quantum information theory. 
	
There are two undoubtedly interesting extensions of this work. The first one is to combine pairs of functions from the present four families into so called $E-$functions, two-variable generalization of the common exponential function are found in \cite{HHP}. In this way we can find six families of functions, which again are orthogonal as continuous and also as discrete functions over finite extension of $F$.
	
The second possible extension of this work is to three-variables orthogonal polynomials. Curiously, the extensive literature about the polynomials contains little information about their discretization. This work opens a way to study these problems.

The paper is organized as follows. Section 2 reviews some basic notations concerning the root systems and Weyl groups. We present the  short and long fundamental domains of the affine Weyl group of $B_3$ and $C_3$. In Section 3 we define four families of orbit functions. In Section 4 we describe in detail orbit functions $S^s-$ and $S^l-$ including their discrete orthogonality and discrete transforms. 

%%%%%%%%%%%%%%%%%%%%%%%%%%%%%%%%%%%%%%%%%%%%%%%%%%%%%%%%%%%
\section{Root systems, affine Weyl groups and fundamental domains}

\subsection{Root systems}\

Consider a simple Lie algebra of rank three and its ordered set of simple roots $\Delta=(\a_1,\a_2,\a_3)$. The set $\Delta$ forms a basis of the three-dimensional real Euclidean space $\R^3$ \cite{Hum,Kane} and satisfies certain specific conditions. There are only two simple Lie algebras of rank three which, with respect to the standard scalar product $\sca{\cdot}{\cdot}$, have two different lenghts of their simple roots --- $B_3$ and $C_3$. The set $\Delta$ is for these two cases decomposed into the set of short simple roots $\Delta_s$ and the set of long simple roots $\Delta_l$:
\begin{equation}\label{decomp}
\Delta=\Delta_s\cup\Delta_l .
\end{equation}
 The set $\Delta$ is usually described by the Coxeter--Dynkin diagram and its corresponding Coxeter matrix $M$ or, equivalently, by the Cartan matrix $C$.  
The vectors called coroots $\a_i^\vee$ are defined as renormalizations of roots: $\a_i^\vee= 2\a_i/\left\langle \a_i, \a_i \right\rangle,\,i=1,\,2,\,3$.
In addition to the $\a-$basis of simple roots and the $\a^\vee-$basis, the following two bases are useful: the weight $\o-$basis, defined by the relations
$$
\left\langle \a_i^\vee,\o_j\right\rangle=\delta_{ij}\,,\quad i,j\in\{1,2,3\}, 
$$ 
and the coweight $\o^\vee-$basis, given by renormalization as $\o_i^\vee= 2\o_i/\left\langle \a_i, \a_i \right\rangle,\,i=1,\,2,\,3.$

Standardly, the root lattice $Q$ is the set of all integer linear combinations of the simple roots
$$
Q = \Z \a_1 + \Z \a_2 +\Z \a_3
$$
and the coroot $Q^\vee$ lattice is 
$$
Q^\vee = \Z \a^\vee_1 + \Z \a^\vee_2+ \Z \a^\vee_3.
$$
The weight lattice and the coweight lattice are given standardly as 
$$
P = \Z \o_1 + \Z \o_2+\Z \o_3,\quad P^\vee = \Z \o^\vee_1 + \Z \o^\vee_2+ \Z \o^\vee_3.
$$
Two important subsets of the weight lattice $P$ are the cone of dominant weights $P^+$ and the cone of strictly dominant weights $P^{++}$:
$$
P^+=\Z^{\geq 0}\o_1+\Z^{\geq 0}\o_2+\Z^{\geq 0}\o_3\  \supset \  P^{++}=\N\o_1+\N\o_2+\N\o_3.
$$

The decomposition \eqref{decomp} induces two subsets of $P^+$ which are crucial for description of the orbit functions. The first excludes points from $P^+$ which are orthogonal to short roots,
\begin{equation}\label{Ps+}
P^{+s}= \set{\o\in P^+}{(\forall \a \in \Delta_s)(\sca{\o}{\a}>0) }
\end{equation}
and the second excludes points orthogonal to long roots
\begin{equation}\label{Pl+}
P^{+l}= \set{\o\in P^+}{(\forall \a \in \Delta_l)(\sca{\o}{\a}>0) }.
\end{equation}

\subsection{Affine Weyl groups}\

The reflection $r_{\a}$, $\a \in \Delta$, which fixes the plane orthogonal to $\a$ and passing through the origin, is explicitly written for $x\in\R^3$ as
$r_{\a}x=x-\langle \a,x\rangle\a^\vee $.
Weyl group $W$ is a finite group generated by reflections $r_i\equiv r_{\a_i},\,i=1,\,2,\,3$. The system of vectors obtained by the action of $W$ on the set of simple roots $\Delta$ forms a root system $W\Delta$ which contains its unique highest root $\xi \in W\Delta$. The marks are the coefficients $m_1,m_2,m_3$ of the highest root $\xi$ in $\a-$basis, $\xi=m_1\a_1 +m_2 \a_2+m_3 \a_3$.

The affine reflection $r_0$ with respect to this highest root is given by
$$
r_0 x=r_\xi x + \frac{2\xi}{\sca{\xi}{\xi}}\,,\qquad
r_{\xi}x=x-\frac{2\sca{x}{\xi} }{\sca{\xi}{\xi}}\xi\,,\qquad x\in\R^3\,.
$$
The affine Weyl group $W^{\mathrm{aff}}$ is generated by reflections from the set $R=\{r_0,r_1,r_2,r_3\}$. 
The decomposition \eqref{decomp} induces a decomposition of the generator set $R$, 
\begin{equation}\label{decompR}
R=R^s\cup R^l
\end{equation} 
where the subsets $R^s$ and $R^l$ are given by
\begin{align*}
R^s&=\set{r_\a}{r_\a\in\Delta_s}\\   
R^l&=\set{r_\a}{r_\a\in\Delta_l}\cup\{r_0\}. 
\end{align*}
The affine Weyl group $W^{\mathrm{aff}}$ consists of orthogonal transformations from $W$ and of shifts by vectors from the coroot lattice $Q^\vee$.
The fundamental domain $F$ of the action of $W^{\mathrm{aff}}$ on $\R^3$ is a tetrahedron with vertices $\left\{ 0, \frac{\o^{\vee}_1}{m_1},\frac{\o^{\vee}_2}{m_2},\frac{\o^{\vee}_3}{m_3} \right\}$.  

The set of reflections corresponding to dual roots $\Delta^\vee=(\a^\vee_1,\a^\vee_2,\a^\vee_3)$ also generates the Weyl group~$W$. The system of vectors $W\Delta^\vee$ is a root system and contains the highest dual root $\eta \in W\Delta^\vee$. The dual marks are the coefficients $m^\vee_1,m^\vee_2,m^\vee_3$ of the dual highest root $\eta$ in $\a^\vee-$basis, $\eta=m^\vee_1\a^\vee_1 +m^\vee_2 \a^\vee_2+m^\vee_3 \a^\vee_3$.

The dual affine reflection $r^\vee_0$ with respect to the highest dual root is given by
$$
r^\vee_0 x=r_\eta x + \frac{2\eta}{\sca{\eta}{\eta}}\,,\qquad
r_{\eta}x=x-\frac{2\sca{x}{\eta} }{\sca{\eta}{\eta}}\eta\,,\qquad x\in\R^3\,.
$$
The dual affine Weyl group $\widehat{W}^{\mathrm{aff}}$ is generated by reflections from the set $R^\vee=\{r^\vee_0,\,r_1,\,r_2,\,r_3\}$, see \cite{HP}. The decomposition \eqref{decomp} also induces a decomposition of the generator set $R^\vee$, 
\begin{equation}\label{decompRd}
R^\vee=\{r_0^\vee,r_1,r_2,r_3\}=R^{s\vee} \cup R^{l\vee}
\end{equation} 
where the subsets $R^{s\vee}$ and $R^{l\vee}$ are given by
\begin{align*}
R^{s\vee}&=\set{r_\a}{r_\a\in\Delta_s}\cup\{r_0^\vee\}\\   
R^{l\vee}&=\set{r_\a}{r_\a\in\Delta_l}.
\end{align*}
The dual affine group $\widehat{W}^{\mathrm{aff}}$ consists of orthogonal transformations from $W$ and of shifts by vectors from the root lattice $Q$.
The dual fundamental domain $F^\vee$ of the action of $\widehat{W}^{\mathrm{aff}}$ on $\R^3$ is a tetrahedron with vertices $\left\{ 0, \frac{\o_1}{m^\vee_1},\frac{\o_2}{m^\vee_2},\frac{\o_3}{m^\vee_3} \right\}$.

\subsection{Short and long fundamental domains}\

The boundary of the fundamental domain $F$ consists of points stabilized by the generators from $R$. Two types of the boundaries of $F$ are determined by the decomposition \eqref{decompR} --- those points which are stabilized by $R^s$ are collected in the short boundary $H^s$,
\begin{equation*}
H^s=\set{a\in F}{(\exists r \in R^s)(ra=a)}
\end{equation*}
and the points stabilized by $R^l$ in the long boundary $H^l$,
\begin{equation*}
H^l=\set{a\in F}{(\exists r \in R^l)(ra=a)}.
\end{equation*}
The points from $F$ which do not lie on the short boundary form the short fundamental domain $F^s$, 
\begin{equation*}
F^s=F\setminus H^s
\end{equation*}
and the points which do not lie on the long boundary form the long fundamental domain $F^l$, 
\begin{equation*}
F^l=F\setminus H^l.
\end{equation*}

Similarly, the boundary of the dual fundamental domain $F^\vee$ consists of points stabilized by the generators from $R^\vee$ and two types of the boundaries of $F^\vee$ are determined by the decomposition \eqref{decompRd}. The points which are stabilized by $R^{s\vee}$ are collected in the short dual boundary $H^{s\vee}$,
\begin{equation*}
H^{s\vee}=\set{a\in F^\vee}{(\exists r \in R^{s\vee})(ra=a)}
\end{equation*}
and the points stabilized by $R^{l\vee}$ in the long dual boundary $H^{l\vee}$,
\begin{equation*}
H^{l\vee}=\set{a\in F^\vee}{(\exists r \in R^{l\vee})(ra=a)}.
\end{equation*}
The points from $F^\vee$ which do not lie on the short dual boundary form the short dual fundamental domain $F^{s\vee}$, 
\begin{equation*}
F^{s\vee}=F^\vee\setminus H^{s\vee}
\end{equation*}
and the points not on the long dual boundary form the long dual fundamental domain $F^{l\vee}$, 
\begin{equation*}
F^{l\vee}=F^\vee\setminus H^{l\vee}.
\end{equation*}

\subsection{The Lie algebra $B_3$}\

\begin{figure}[ht!]
\begin{center}
\includegraphics[width=3in]{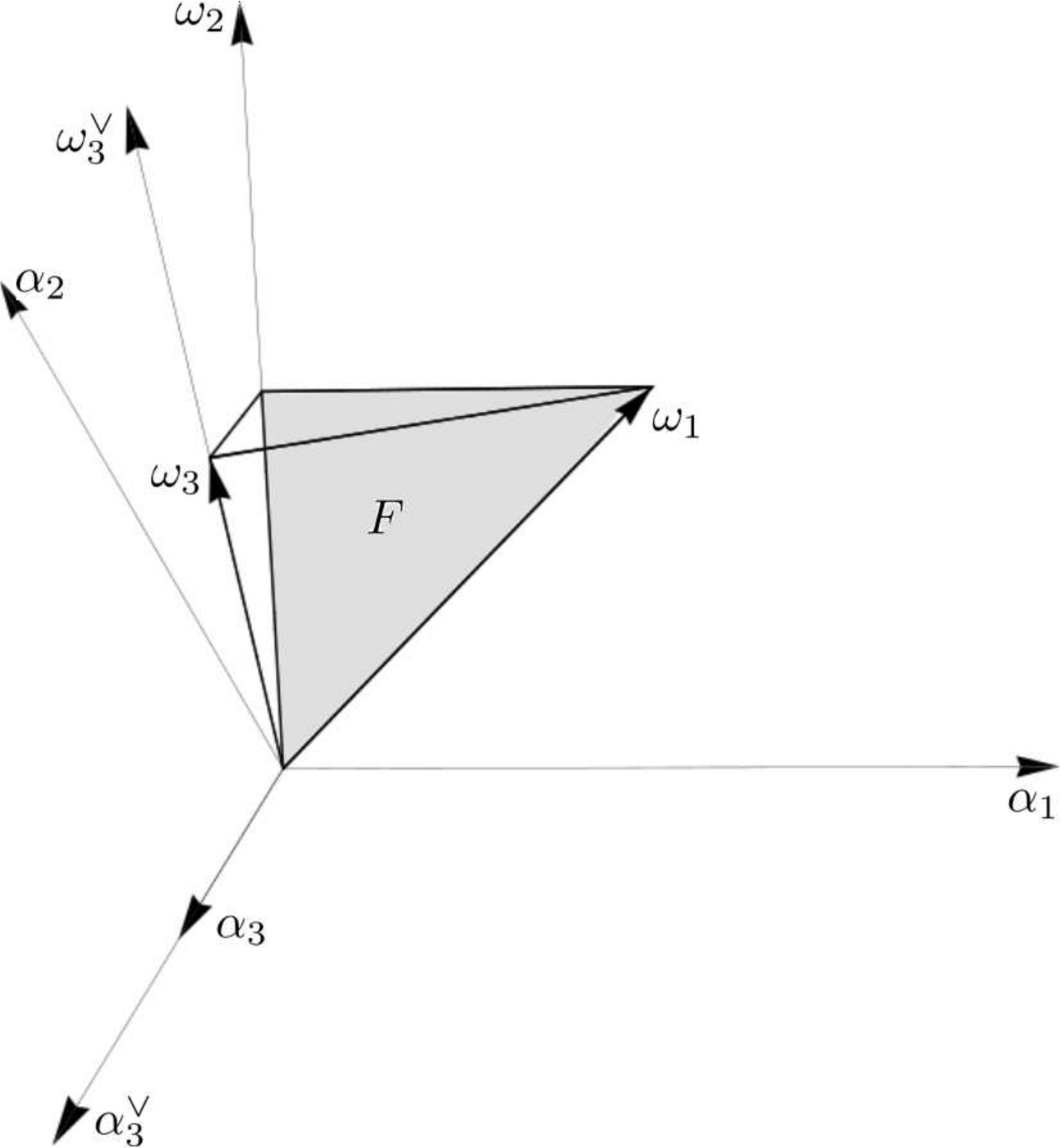}
\caption{The $\a$ and $\o-$bases and the fundamental domain $F$ of $B_3$. The tetrahedron $F$ without the grey back face, which depicts $H^s$, is the short fundamental domain $F^s$; the tetrahedron $F$ without the three unmarked faces is the long fundamental domain $F^l$.}\label{figB3}
\end{center}
\end{figure}

For practical purposes, the most convenient way of specifying the root system $\Delta$ and the bases $\a^\vee$, $\o^\vee$ and $\o$ is to evaluate their coordinates in a fixed orthonormal basis. With respect to the standard orthonormal basis of $\R^3$, these four bases of $B_3$ are of the form 
\begin{alignat*}{4}
\a_1&=(1,-1,0), &\qquad \o_1&=(1,0,0),         &\qquad \a^\vee_1&=(1,-1,0),  &\qquad  \o^\vee_1&=(1,0,0),\\
\a_2&=(0,1,-1), &\qquad \o_2&=(1,1,0),     &\qquad \a^\vee_2&=(0,1,-1),  &\qquad  \o^\vee_2&=(1,1,0),\\
\a_3&=(0,0,1),    &\qquad \o_3&=(\tfrac12,\tfrac12,\tfrac12), &\qquad \a^\vee_3&=(0,0,2),       &\qquad  \o^\vee_3&=(1,1,1).
\end{alignat*}
In this setting it holds that $\sca{\a_1}{\a_1}=\sca{\a_2}{\a_2}=2$ and $\sca{\a_3}{\a_3}=1$, which means that $\a_1$, $\a_2$ are the long roots and $\a_3$ is the short root of $B_3$ --- the decomposition \eqref{decomp} is 
\begin{equation*} 
\Delta_s=\{\a_3\},\qquad \Delta_l=\{\a_1,\a_2\} .
\end{equation*}
The short and long subsets \eqref{Ps+}, \eqref{Pl+} of the grid $P^+$ are of the form
\begin{equation*} 
P^{+s}= \Z^{\geq 0}\o_1 + \Z^{\geq 0}\o_2 + \N\o_3,\qquad  P^{+l}= \N\o_1 + \N\o_2 + \Z^{\geq 0}\o_3.  
\end{equation*}
The highest root $\xi$ and the dual highest root $\eta$ are given as
\begin{equation*} 
\xi= \a_1+2\a_2+2\a_3,\qquad \eta= 2\a^\vee_1+2\a^\vee_2+\a^\vee_3
\end{equation*}
which determine the fundamental domain $F$ and the dual fundamental domain $F^\vee$ explicitly as
\begin{align*}
F&= \set{y_1\o_1^\vee+y_2\o_2^\vee+y_3\o_3^\vee}{y_0,\,y_1,\,y_2,\,y_3\in \R^{\geq 0},\,y_0+y_1+2y_2+2y_3=1},\\
F^\vee&= \set{z_1\o_1+z_2\o_2+z_3\o_3}{z_0,\,z_1,\,z_2,\,z_3\in \R^{\geq 0},\,z_0+2z_1+2z_2+z_3=1}.
\end{align*}
The induced decompositions \eqref{decompR}, \eqref{decompRd} are of the form
\begin{alignat*}{3}
R^s&= \{r_3 \},&\qquad  R^l&=\{r_0,\,r_1,\,r_2 \},\\  
R^{s\vee}&= \{r_0^\vee,\,r_3 \},&\qquad  R^{l\vee}&= \{r_1,\,r_2 \} ,
\end{alignat*}
which give the short and the long fundamental domains explicitly
\begin{align*}
F^s&=\set{y_1^s\o_1^\vee+y_2^s\o_2^\vee+y_3^s\o_3^\vee}{y^s_0,\,y^s_1,\,y^s_2\in \R^{\geq 0},\,y^s_3\in \R^{> 0},\, y_0^s+y_1^s+2y_2^s+2y_3^s=1},\\   
F^l&=\set{y_1^l\o_1^\vee+y_2^l\o_2^\vee+y_3^l\o_3^\vee}{y^l_0,\,y^l_1,\,y^l_2\in \R^{> 0},\,y^l_3\in \R^{\geq 0},\,y_0^l+y_1^l+2y_2^l+2y_3^l=1},
\end{align*}
together with their dual versions
\begin{align*}
F^{s\vee}&=\set{z_1^s\o_1+z_2^s\o_2+z_3^s\o_3}{z_0^s,\,z_3^s\in\R^{> 0},\,z_1^s,\,z_2^s\in\R^{\geq 0},\,z_0^s+2z_1^s+2z_2^s+z_3^s=1},\\   
F^{l\vee}&=\set{z_1^l\o_1+z_2^l\o_2+z_3^l\o_3}{z_0^l,\,z_3^l\in\R^{\geq 0},\,z_1^l,\,z_2^l\in\R^{> 0},\,z_0^l+2z_1^l+2z_2^l+z_3^l=1}.
\end{align*}
The $\a$ and $\o-$bases, together with the fundamental domains $F$, $F^s$ and $F^l$ of $B_3$ are depicted in Figure~\ref{figB3}. 

\subsection{The Lie algebra $C_3$}\

\begin{figure}[!ht]
\begin{center}
\includegraphics[width=3in]{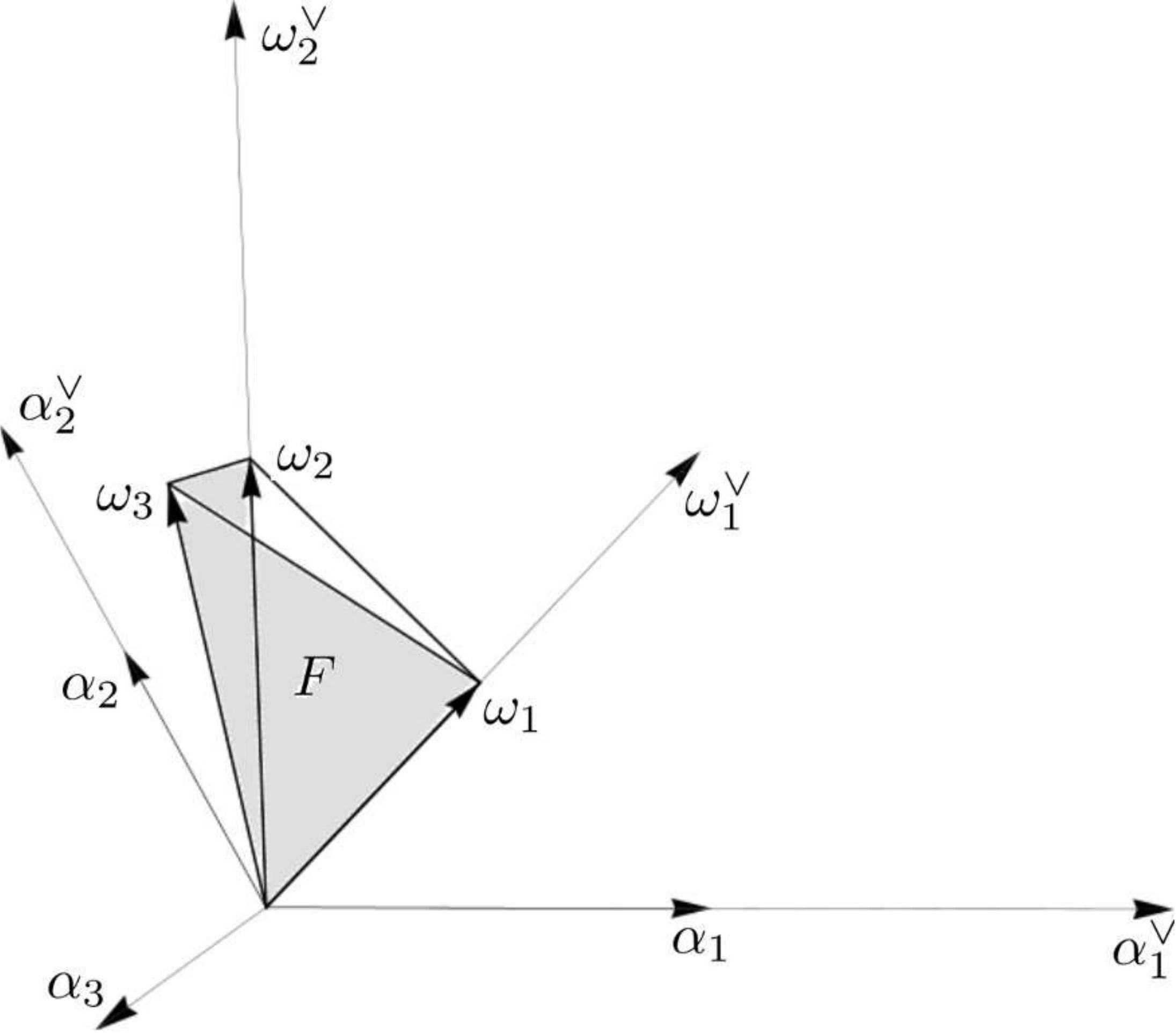}
\caption{ The $\a$ and $\o-$bases and the fundamental domain $F$ of $C_3$. The tetrahedron $F$ without the two grey faces, which depict $H^s$, is the short fundamental domain $F^s$; the tetrahedron $F$ without the two unmarked faces is the long fundamental domain $F^l$.}\label{figC3}
\end{center}
\end{figure}

With respect to the standard orthonormal basis of $\R^3$, the four bases of $C_3$ are of the form 
\begin{alignat*}{4}
\a_1&=\frac{1}{\sqrt{2}}(1,-1,0), &\qquad \o_1&=\frac{1}{\sqrt{2}}(1,0,0),         &\qquad \a^\vee_1&=\sqrt{2}\,(1,-1,0),  &\qquad  \o^\vee_1&=(\sqrt{2},0,0),\\
\a_2&=\frac{1}{\sqrt{2}}(0,1,-1), &\qquad \o_2&=\frac{1}{\sqrt{2}}(1,1,0),     &\qquad \a^\vee_2&=\sqrt{2}\,(0,1,-1),  &\qquad  \o^\vee_2&=\sqrt{2}\,(1,1,0),\\
\a_3&=(0,0,\sqrt{2}),    &\qquad \o_3&=\frac{1}{\sqrt{2}}(1,1,1), &\qquad \a^\vee_3&=(0,0,\sqrt{2}),       &\qquad  \o^\vee_3&=\frac{1}{\sqrt{2}}(1,1,1).
\end{alignat*}
In this setting it holds that $\sca{\a_1}{\a_1}=\sca{\a_2}{\a_2}=1$ and $\sca{\a_3}{\a_3}=2$, which means that $\a_1$, $\a_2$ are the short roots and $\a_3$ is the long root of $C_3$ --- the decomposition \eqref{decomp} is 
\begin{equation*} 
\Delta_s=\{\a_1,\a_2\},\qquad \Delta_l=\{\a_3\} .
\end{equation*}
The short and long subsets \eqref{Ps+}, \eqref{Pl+} of the grid $P^+$ are of the form
\begin{equation*} 
P^{+s}= \N\o_1 + \N\o_2 + \Z^{\geq 0}\o_3,\qquad  P^{+l}= \Z^{\geq 0}\o_1 + \Z^{\geq 0}\o_2 + \N\o_3.  
\end{equation*}
The highest root $\xi$ and the dual highest root $\eta$ are given as
\begin{equation*} 
\xi= 2\a_1+2\a_2+\a_3,\qquad \eta= \a^\vee_1+2\a^\vee_2+2\a^\vee_3
\end{equation*}
which determine the fundamental domain $F$ and the dual fundamental domain $F^\vee$ explicitly as
\begin{align*}
F&= \set{y_1\o_1^\vee+y_2\o_2^\vee+y_3\o_3^\vee}{y_0,\,y_1,\,y_2,\,y_3\in \R^{\geq 0},\,y_0+2y_1+2y_2+y_3=1},\\
F^\vee&= \set{z_1\o_1+z_2\o_2+z_3\o_3}{z_0,\,z_1,\,z_2,\,z_3\in \R^{\geq 0},\,z_0+z_1+2z_2+2z_3=1}.
\end{align*}
The induced decompositions \eqref{decompR}, \eqref{decompRd} are of the form
\begin{alignat*}{3}
R^s&=\{r_1,\,r_2 \},&\qquad  R^l&=\{r_0,\,r_3 \},\\  
R^{s\vee}&= \{r_0^\vee,\,r_1,\,r_2 \},&\qquad  R^{l\vee}&=\{r_3 \} ,
\end{alignat*}
which give the short and the long fundamental domains explicitly
\begin{align*}
F^s&=\set{y_1^s\o_1^\vee+y_2^s\o_2^\vee+y_3^s\o_3^\vee}{y^s_0,\,y^s_3\in \R^{\geq 0},\,y^s_1,\,y^s_2\in \R^{> 0},\, y_0^s+2y_1^s+2y_2^s+y_3^s=1},\\   
F^l&=\set{y_1^l\o_1^\vee+y_2^l\o_2^\vee+y_3^l\o_3^\vee}{y^l_0,\,y^l_3\in \R^{> 0},\,y^l_1,\,y^l_2\in \R^{\geq 0},\,y_0^l+2y_1^l+2y_2^l+y_3^l=1},
\end{align*}
together with their dual versions
\begin{align*}
F^{s\vee}&=\set{z_1^s\o_1+z_2^s\o_2+z_3^s\o_3}{z_0^s,\,z_1^s,\,z_2^s\in\R^{> 0},\,z_3^s\in\R^{\geq 0},\,z_0^s+z_1^s+2z_2^s+2z_3^s=1},\\   
F^{l\vee}&=\set{z_1^l\o_1+z_2^l\o_2+z_3^l\o_3}{z_0^l,\,z_1^l,\,z_2^l\in\R^{\geq 0},\,z_3^l\in\R^{> 0},\,z_0^l+z_1^l+2z_2^l+2z_3^l=1}.
\end{align*}
The $\a$ and $\o-$bases, together with the fundamental domains $F$, $F^s$ and $F^l$ of $C_3$ are depicted in Figure~\ref{figC3}. 

\section{Orbit functions}

\subsection{Orbits and stabilizers}\

Considering any $\la\in \R^3$, the stabilizer $\mathrm{Stab}_W (\la)$ of $\la$ is the set
$\mathrm{Stab}_W(\la)=\set{w\in W}{w\la=\la}$
and its order is denoted by $d_{\la}$,
\begin{equation}\label{dla}
d_{\la}\equiv  |\mathrm{Stab}_{W} (\la)|.
\end{equation}

For calculation of continuous orthogonality of various types of orbit functions, the number of elements in the Weyl group $W$  and the volume of the fundamental domain $F$ are needed. Their product $|W||F|$ is denoted by $K$ and it holds that (see e.g. \cite{HP})
\begin{equation}\label{K}
K\equiv |W||F|= 
  \begin{cases}
    2 & \text{for } B_3 \\
      2\sqrt{2}    & \text{for } C_3.
  \end{cases}
\end{equation}

The orbits and the stabilizers on the torus $\R^3/Q^{\vee}$ are needed for the discrete calculus of orbit functions. An arbitrarily chosen natural number $M$ controls the density of the grids appearing in this calculus~\cite{HP}. The discrete calculus of orbit functions is performed over the finite group $\frac{1}{M}P^{\vee}/Q^{\vee}$. The finite complement set of weights is taken as the quotient group $P/MQ$.  
For any $x\in \R^3/Q^{\vee}$, its orbit by the action of $W$ is given by
$W x=\set{wx\in \R^3/Q^{\vee} }{w\in W}$
and its order is denoted by $\ep (x)$,
\begin{equation}\label{epx}
\ep (x)\equiv |W x |.
\end{equation}
For any $\la\in P/MQ$, its stabilizer $\mathrm{Stab}^{\vee} (\la)$ by the action of $W$ is given by
$$\mathrm{Stab}^{\vee} (\la)=\set{w\in W}{w\la=\la}.$$
and its order is denoted by $h^{\vee}_{\la}$
\begin{equation}\label{hla}
h^{\vee}_{\la}\equiv |\mathrm{Stab}^{\vee} (\la)|.
\end{equation}

Moreover, for calculation of discrete orthogonality of various types of orbit functions, the determinant of the Cartan matrix $\det C$ is needed. The product $|W|\det C$ is denoted by $k$ and it holds that (see e.g.~\cite{HP})
\begin{equation}\label{k}
k\equiv |W|\det C= 96, \qquad \text{for } B_3,\, C_3.
\end{equation}

\subsection{Four types of orbit functions}\

The Weyl group $W$ can also be abstractly defined by the presentation of a Coxeter group 
\begin{equation}\label{presentation}
r_i^2=1,\quad (r_ir_j)^{m_{ij}}=1,\quad i,j=1,\,2,\,3,
\end{equation}
where integers $m_{ij}$ denote elements of the Coxeter matrix. 
The Coxeter matrices $M$ of $B_3$ and $C_3$ are of the form
\begin{equation}\label{Coxmat}
M(B_3)=\begin{pmatrix}  1&3&2\\3&1&4\\2&4&1\end{pmatrix}\,,\qquad M(C_3)=\,\begin{pmatrix} 1&4&2\\4&1&3\\2&3&1\\\end{pmatrix}\,.
\end{equation}

Crucial tool for defining the orbit functions are 'sign' homomorphisms $\sigma:W\rightarrow\{\pm1 \}$. A sign homomorphism can be defined by prescribing its values on the generators $r_1,\,r_2,\,r_3$ of $W$. An admissible mapping has to satisfy the presentation condition \eqref{presentation}
\begin{equation}\label{admit}
\sigma(r_i)^2=1,\quad (\sigma(r_i)\sigma(r_j))^{m_{ij}}=1,\quad i,j=1,2,3.
\end{equation}
Two obvious choices $\sigma(r_i)=1$ and $\sigma^e(r_i)=-1$ for every $i\in \{1,2,3\}$ lead to the standard homomorphisms $\id$ and $\sigma^e$ with values given for any $w\in W$ as
$$\begin{aligned}
\id(w)&=1, \label{homid} \\
\sigma^e(w)&=\det w. \label{home}
\end{aligned}$$
It turns out that for root systems with two different lengths of roots there are two other available choices~\cite{MMP}. This can also be directly seen for the cases of $B_3$ and $C_3$ ---
the non-diagonal elements $m_{ij}$ of the Coxeter matrices (\ref{Coxmat}) are even except for the elements corresponding simultaneously to two short roots or two long roots. Therefore, if we set one value of $\sigma$ on all the short roots and, independently, another value for all the long roots, the admissibility condition \eqref{admit} is still satisfied. Consequently, there are two more sign homomorphisms, denoted by $\sigma^s$ and $\sigma^l$, 
\begin{align*}
\sigma^s(r_\a)&=\begin{cases} 1,\quad \a\in \Delta_l  \\ -1,\quad \a\in \Delta_s\end{cases}\\
\sigma^l(r_\a)&=\begin{cases} 1,\quad \a\in \Delta_s  \\ -1,\quad \a\in \Delta_l.\end{cases}
\end{align*}

Each of the sign homomorphisms $\id$, $\sigma^e$, $\sigma^s$ and $\sigma^l$ induces a family of complex orbit functions. The functions in each family are labeled by the weights $\la\in P$ and their general form is
\begin{equation}\label{genorb}
\psi^\sigma_{\la}(a)=\sum_{w\in W}\sigma (w)\, e^{2 \pi i \sca{ w\la}{a}},\quad a\in \R^3.
\end{equation}
Among the basic general properties of all these families of orbit functions are their invariance with respect to shifts from $q^{\vee} \in Q^{\vee}$
\begin{equation}\label{genshift}
\psi^\sigma_{\la}(a+q^\vee)=\psi^\sigma_{\la}(a)
\end{equation}
and their invariance or antiinvariance with respect to action of elements from $w\in W$
\begin{align}
\psi^\sigma_{\la}(wa)&=\sigma (w)\,\psi^\sigma_{\la}(a) \label{geninv1}\\
\psi^\sigma_{w\la}(a)&=\sigma (w)\,\psi^\sigma_{\la}(a).\label{geninv2}
\end{align}

\subsection{$C-$ and $S-$functions}\

Choosing $\sigma=\id$, so-called $C-$functions \cite{KP2} are obtained from \eqref{genorb}. Following \cite{HP}, these functions are here denoted  by the symbol $\Phi_\la\equiv \psi^{\id}_\la$. The invariance \eqref{genshift}, \eqref{geninv1} with respect to the affine Weyl group $W^{\mathrm{aff}}$ allows to consider $\Phi_\la$ only on the fundamental domain $F$.
Similarly, the invariance \eqref{geninv2} restricts the weights $\la\in P$ to the set $P^+$.
Thus, we have
\begin{equation*}%\label{orbe+}
\Phi_\la(x)=\sum_{w\in W} e^{2 \pi i \sca{ w\la}{x}},\qquad x\in F,\, \la \in P^+.
\end{equation*} 
For the detailed review of $C-$functions see \cite{KP2} and references therein. Continuous and discrete orthogonality of  $C-$functions $\Phi_\la$ are studied for all rank two cases in detail in \cite{PZ1,PZ2}. The discretization properties of 
$C-$functions on a finite fragment
of the grid $\frac{1}{M}P^{\vee}$ is described in full generality in \cite{HP}.

Choosing $\sigma=\sigma^e$, well--known $S-$functions \cite{KP3} are obtained from \eqref{genorb}. Following \cite{HP,KP3}, these functions are here denoted  by the symbol $\varphi_\la\equiv \psi^{\sigma^e}_\la$. The invariance \eqref{genshift} together with the antiinvariance \eqref{geninv1}  allows to consider $\varphi_\la$ only on the interior $F^\circ$ of the fundamental domain $F$.
Similarly, the antiinvariance \eqref{geninv2} restricts the weights $\la\in P$ to the set $P^{++}$.
Thus, we have
\begin{equation*}
\varphi_\la(x)=\sum_{w\in W} (\det w)  \, e^{2 \pi i \sca{ w\la}{x}},\qquad x\in F^\circ,\, \la \in P^{++}.
\end{equation*} 
For the detailed review of $S-$functions see \cite{KP3} and references therein. Continuous and discrete orthogonality of  $S-$functions $\varphi_\la$ are studied for all rank two cases in detail in \cite{PZ3}. The discretization properties of 
$S-$functions on a finite fragment
of the grid $\frac{1}{M}P^{\vee}$ is described in full generality in \cite{HP}.

\section{$S^s-$ and $S^l-$functions}

\subsection{$S^s-$functions}\

Choosing $\sigma=\sigma^s$, so--called $S^s-$functions \cite{MMP} are obtained from \eqref{genorb}. Following \cite{HMP}, these functions are here denoted by the symbol $\varphi^s_\la\equiv \psi^{\sigma^s}_\la$. The antiinvariance \eqref{geninv1} with respect to the short reflections $r_\a,\, \a\in \Delta_s$ together with shift invariance \eqref{genshift}
imply zero values of $S^s-$functions on the boundary $H^s$,
$$\varphi^s_{\la}(a')=0,\quad  a'\in H^s.$$
Therefore, the functions $\varphi^s_{\la}$ are considered on the fundamental domain $F^s=F\setminus H^s$ only.
Similarly, the antiinvariance \eqref{geninv2} restricts the weights $\la\in P$ to the set $P^{+s}$.
Thus, we have
\begin{equation*}
\varphi^s_\la(x)=\sum_{w\in W} \sigma^s (w)  \, e^{2 \pi i \sca{ w\la}{x}},\qquad x\in F^s,\, \la \in P^{+s}.
\end{equation*} 

\subsubsection{Continuous orthogonality and $S^{s}-$transforms}
For any two weights $\la,\la'\in P^{+s}$ the
corresponding $S^{s}-$functions are orthogonal on $F^{s}$
\begin{equation}\label{scorthog}
\int_{{F}^{s}}\varphi^{s}_{\lambda}(x)\overline{\varphi^{s}_{\lambda'}(x)}\,\mathrm{d}x=
K\,d_{\la}\,\delta_{\lambda\lambda'} \end{equation}
where $d_{\la}$, $K$ are given by (\ref{dla}), (\ref{K}), respectively. The $S^s-$functions determine symmetrized Fourier series
expansions,
\begin{equation*}
f(x)=\sum_{\la\in P^{+s}}c^{s}_{\la}\varphi^{s}_{\la}(x),\quad {\mathrm{
where}}\
c^{s}_{\la}=\frac{1}{K d_{\la}}\int_{F^s}f(x)\overline{\varphi^{s}_{\la}(x)}\,\mathrm{d}x.
\end{equation*}

\subsubsection{Discrete orthogonality and discrete $S^{s}-$transforms}\ The finite set of points is given by $$F_M^{s}=\frac{1}{M}P^\vee / Q^\vee \cap F^{s}$$
and the corresponding finite set of weights as 
$$\Lambda^{s}_M = P/MQ \cap MF^{s\vee}.$$
Then, for $\la,\la' \in\Lambda^{s}_M$, the following discrete orthogonality relations hold,
\begin{equation}\label{sdortho}
 \sum_{x\in F^{s}_M}\ep(x) \varphi^{s}_\la(x)\overline{\varphi^{s}_{\la'}(x)}=kM^3 h^{\vee}_\la \delta_{\la\la'}
\end{equation}
where $\ep(x)$, $h^{\vee}_{\la}$ and $k$ are given by \eqref{epx}, \eqref{hla} and \eqref{k}, respectively. The discrete symmetrized $S^s-$function expansion is given by       \begin{equation}\label{disc_transform}
f(x)=\sum_{\la\in \Lambda_M^{s}}c^{s}_{\la}\varphi^{s}_{\la}(x),\quad {\mathrm{
where}}\
c^{s}_{\la}=\frac{1}{k M^3 h^{\vee}_\la}\sum_{x\in F^s_M}\ep(x) f(x)\overline{\varphi^{s}_{\la}(x)}.
\end{equation}

\subsubsection{$S^s-$functions of $B_3$}
\begin{figure}
\begin{center}
\includegraphics[width=5in]{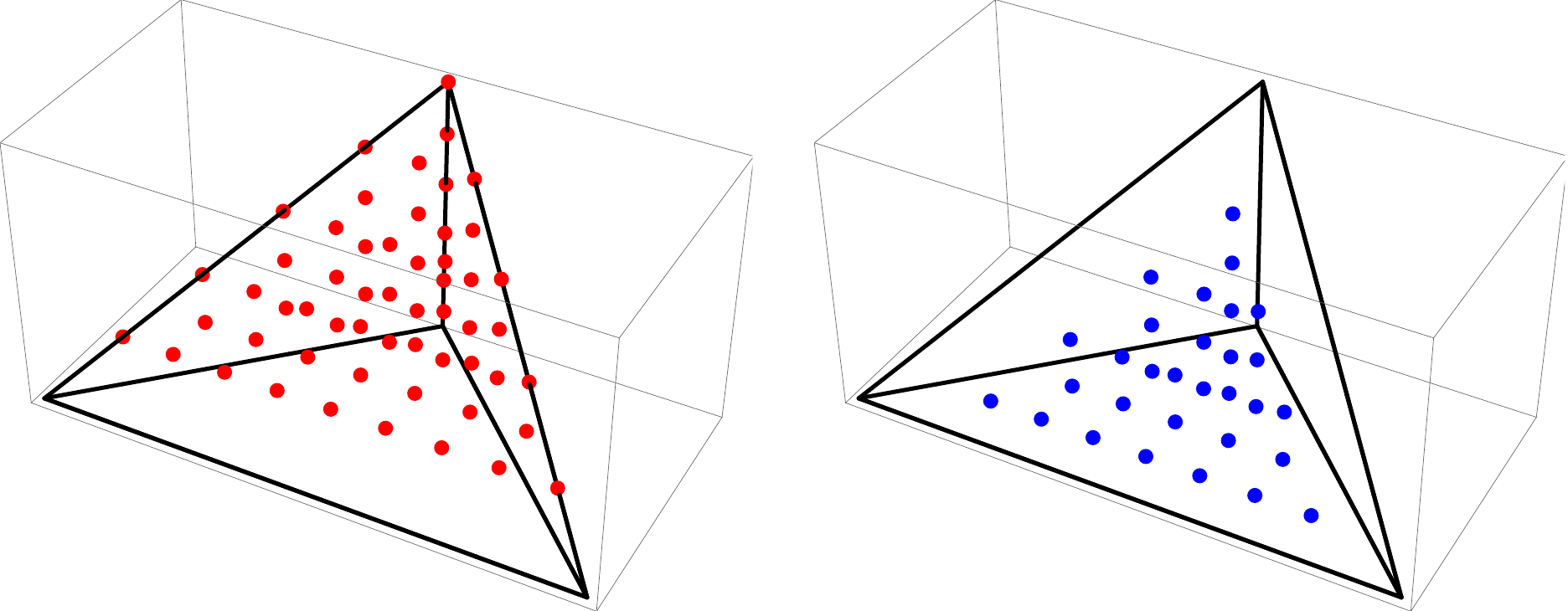}
\caption{The grids $F_{10}^s$ and $F_{10}^l$ of $B_3$. On the left--hand side are depicted the points of the grid $F_{10}^s$, given by \eqref{FMB3}; on the right--hand side are depicted the points of $F_{10}^l$, given by \eqref{FMB3l}.}\label{gridB3}
\end{center}
\end{figure}

For a point with coordinates in $\alpha^\vee-$basis $(x,y,z)$ and a weight with coordinates in $\o-$basis of $(a,b,c)$, the coressponding $S^s-$functions are explicitly evaluated as
{\small\begin{align*}
\varphi^s_{\la}&(x,y,z)
=2i\left\lbrace \sin(2\pi(ax+by+cz)) + \sin(2\pi(-ax+(a+b)y+cz))\right.\\  
& \left.+ \sin(2\pi((a+b)x-by+(2b+c)z)) - \sin(2\pi(ax+(b+c)y-cz))\right.\\ 
& \left.+ \sin(2\pi(bx-(a+b)y+(2a+2b+c)z)) - \sin(2\pi(-ax+(a+b+c)y-cz))\right.\\
& \left.+ \sin(2\pi(-(a+b)x+ay+(2b+c)z)) - \sin(2\pi((a+b)x+(b+c)y-(2b+c)z)) \right.\\
& \left.- \sin(2\pi((a+b+c)x-(b+c)y+(2b+c)z)) + \sin(2\pi(-bx-ay-(2a+2b+c)z)) \right.\\
& \left.- \sin(2\pi(bx+(a+b+c)y-(2a+2b+c)z)) - \sin(2\pi((b+c)x-(a+b+c)y+(2a+2b+c)z)) \right.\\
& \left.- \sin(2\pi(-(a+b)x+(a+2b+c)y-(2b+c)z)) - \sin(2\pi((a+2b+c)x-(b+c)y+cz)) \right.\\
& \left.- \sin(2\pi(-(a+b+c)x+ay+(2b+c)z)) + \sin(2\pi((a+b+c)x+by-(2b+c)z)) \right.\\
& \left.- \sin(2\pi(-bx+(a+2b+c)y-(2a+2b+c)z)) - \sin(2\pi((a+2b+c)x-(a+b+c)y+cz) \right.\\
& \left.- \sin(2\pi(-(b+c)x-ay+(2a+2b+c)z)) + \sin(2\pi((b+c)x+(a+b)y-(2a+2b+c)z) \right.\\
& \left.- \sin(2\pi((b+c)x-(a+2b+c)y+(2a+2b+c)z)) - \sin(2\pi(-(a+2b+c)x+(a+b)y+cz) \right.\\
& \left.+ \sin(2\pi((a+2b+c)x-by-cz)) + \sin(2\pi(-(a+b+c)x+(a+2b+c)y-(2b+c)z) \right\rbrace .
\end{align*}}
The coefficients $d_\la$ of continuous orthogonality relations \eqref{scorthog} are given in Table \ref{tabStab}.
{\small\begin{table}
\begin{tabular}{|c|c||c|c|}
\hline
$\la\in P^+$ & $d_\la$ & $\la\in P^+$ & $d_\la$\\
\hline\hline
$(a,b,c)$ & 1 &
$(a,b,0)$ & 2\\
$(a,0,c)$ & 2&
$(0,b,c)$ & 2\\ 
$(a,0,0)$ & 8&
$(0,b,0)$ & 4\\
$(0,0,c)$ & 6&
$(0,0,0)$ & 48\\ 
\hline
\end{tabular}
\medskip
\caption{Orders of stabilizers of $\la\in P^+$ for $B_3$ and $C_3$, The coordinates $(a,b,c)$ are in $\o-$basis with $a,b,c\neq 0$.}\label{tabStab}
\end{table}}
The discrete grid $F_M^s$ is given by
\begin{equation}\label{FMB3}
F^s_M=\set{\dfrac{u_1^s}{M}\o_1^\vee+\dfrac{u_2^s}{M}\o_2^\vee +\dfrac{u_3^s}{M}\o_3^\vee}{u_0^s,u_1^s,u_2^s\in\Z^{\ge0},u_3^s\in\N,u_0^s+u_1^s+2u_2^s+2u_3^s=M }
\end{equation}
and the corresponding finite set of weights has the form
\begin{align}\label{LMB3}\Lambda_M^s&=\set{t_1^s\o_1+t_2^s\o_2+t_3^s\o_3}{t_0^s,t_3^s\in \N,t_1^s,t_2^s\in\Z^{\ge0},t_0^s+2t_1^s+2t_2^s+t_3^s=M}.
\end{align}
The number of points in each of these grids are given as
\begin{align*}
|F_{2k}^s|&=|\Lambda_{2k}^s|=\dfrac{1}{6}k(k+1)(2k+1) \\
|F_{2k+1}^s|&=|\Lambda_{2k+1}^s|=\dfrac{1}{3}k(k+1)(k+2). 
\end{align*}
The grid $F^s_{10}$ of $B_3$ is depicted in Figure \ref{gridB3}.
The coefficients $\ep(x)$ and $h_\la^{\vee }$ of discrete orthogonality relations (\ref{sdortho}) are given in Table \ref{epsilonyB3}. Each point $x\in F^s_M$ and each weight $\lambda\in \Lambda^s_M$ are represented by the coordinates $[u^s_0,\,u^s_1,\,u^s_2,\,u^s_3]$ and $[t^s_0,\,t^s_1,\,t^s_2,\,t^s_3]$ from defining equations \eqref{FMB3}, \eqref{LMB3}, respectively.  
{\small\begin{table}
\begin{tabular}{c|c}
$x\in F^s_M$ & $\ep(x)$  \\ \hline
$[u^s_0,u^s_1,u^s_2,u^s_3]$ & $48$  \\
$[0,u^s_1,u^s_2,u^s_3]$ & $24$   \\
$[u^s_0,0,u^s_2,u^s_3]$ & $24$   \\
$[u^s_0,u^s_1,0,u^s_3]$ & $24$  \\
$[0,0,u^s_2,u^s_3]$ & $12$   \\
$[u^s_0,0,0,u^s_3]$ & $8$   \\
$[0,u^s_1,0,u^s_3]$ & $8$  \\
$[0,0,0,u^s_3]$ & $2$  \\
\multicolumn{1}{c}{}&\\
$x\in F^l_M$ & $\ep(x)$  \\ \hline
$[u^l_0,u^l_1,u^l_2,u^l_3]$ & $48$  \\
$[u^l_0,u^l_1,u^l_2,0]$ & $24$   \\
\end{tabular}\hspace{12pt}
\begin{tabular}{c|c}
$\la\in \Lambda^s_M$  & $h_\lambda ^{\vee}$ \\ \hline
$[t^s_0,t^s_1,t^s_2,t^s_3]$ & $1$  \\
$[t^s_0,0,t^s_2,t^s_3]$ & $2$   \\
$[t^s_0,t^s_1,0,t^s_3]$ & $2$  \\
$[t^s_0,0,0,t^s_3]$ & $6$   \\
\multicolumn{1}{c}{}&\\
\multicolumn{1}{c}{}&\\
\multicolumn{1}{c}{}&\\
$\la\in \Lambda^l_M$  & $h_\lambda ^{\vee}$ \\ \hline
$[t^l_0,t^l_1,t^l_2,t^l_3]$ & $1$  \\
$[0,t^l_1,t^l_2,t^l_3]$ & $2$   \\
$[t^l_0,t^l_1,t^l_2,0]$ & $2$  \\
$[0,t^l_1,t^l_2,0]$ & $4$  \\
\end{tabular}\\
\medskip
\caption{The coefficients $\ep(x)$ and $h_\lambda ^{\vee}$ of $B_3$. All variables $u^s_i, t^s_i$ and $u^l_i, t^l_i$, $i=0,1,2,3$, are assumed to be natural numbers.}\label{epsilonyB3}
\end{table}}

\subsubsection{$S^s-$functions of $C_3$}
\begin{figure}
\begin{center}
\includegraphics[width=4.5in]{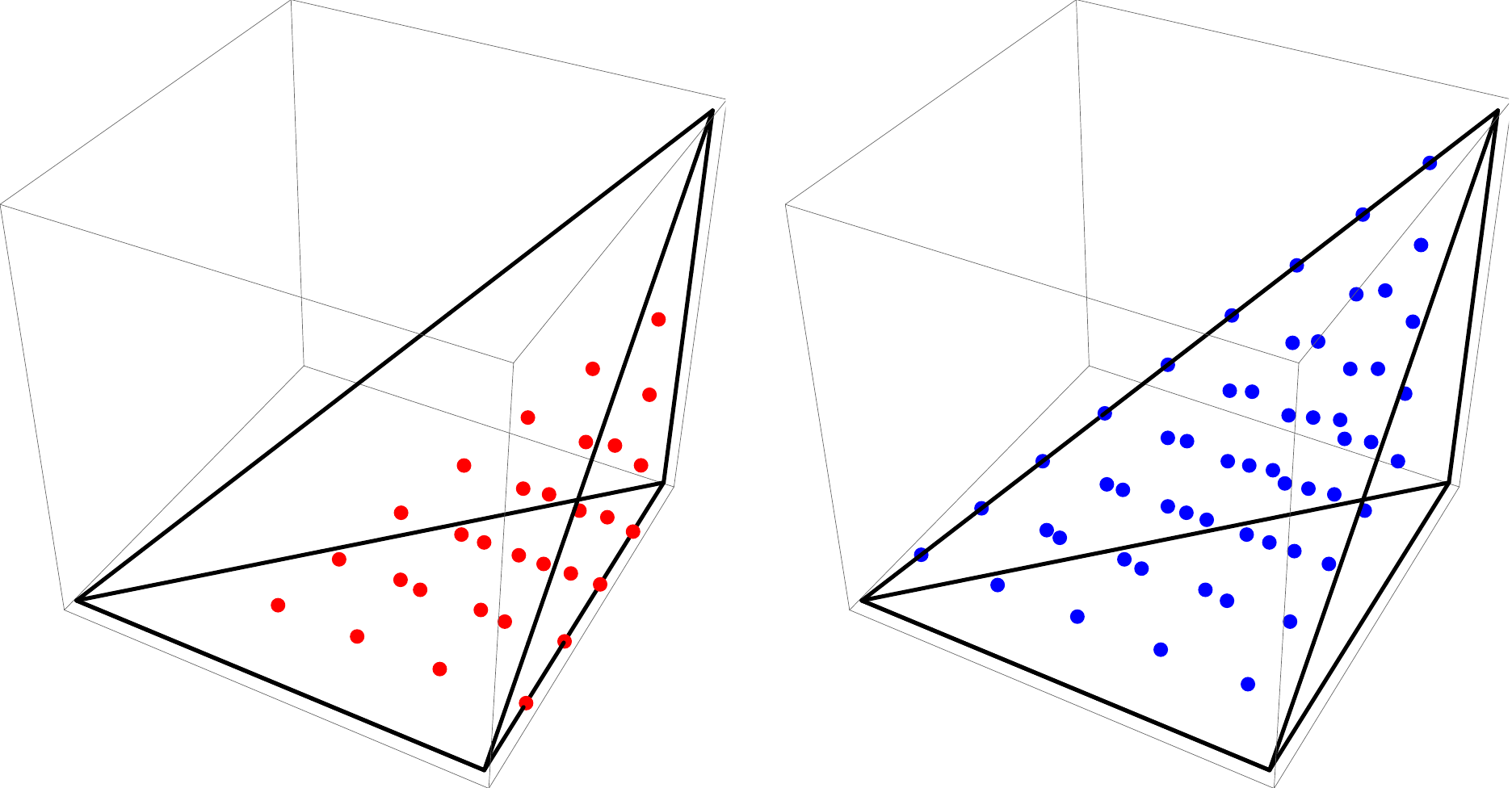}
\caption{The grids $F_{10}^s$ and $F_{10}^l$ of $C_3$. On the left--hand side are depicted the points of the grid $F_{10}^s$, given by \eqref{FMC3}; on the right--hand side are depicted the points of $F_{10}^l$, given by \eqref{FMC3l}.}\label{gridC3}
\end{center}
\end{figure}

For a point with coordinates in $\alpha^\vee-$basis $(x,y,z)$ and a weight with coordinates in $\o-$basis of $(a,b,c)$, the coressponding $S^s-$functions are explicitly evaluated as
{\small\begin{align*}
\varphi^s_{\la}&(x,y,z)
=2\left\lbrace \cos(2\pi(ax+by+cz)) - \cos(2\pi(-ax+(a+b)y+cz))\right.\\  
& \left.- \cos(2\pi((a+b)x-by+(b+c)z)) + \cos(2\pi(ax+(b+2c)y-cz))\right.\\ 
& \left.+ \cos(2\pi(bx-(a+b)y+(a+b+c)z)) - \cos(2\pi(-ax+(a+b+2c)y-cz))\right.\\
& \left.+ \cos(2\pi(-(a+b)x+ay+(b+c)z)) - \cos(2\pi((a+b)x+(b+2c)y-(b+c)z)) \right.\\
& \left.- \cos(2\pi((a+b+2c)x-(b+2c)y+(b+c)z)) - \cos(2\pi(-bx-ay-(a+b+c)z)) \right.\\
& \left.+ \cos(2\pi(bx+(a+b+2c)y-(a+b+c)z)) + \cos(2\pi((b+2c)x-(a+b+2c)y+(a+b+c)z)) \right.\\
& \left.+ \cos(2\pi(-(a+b)x+(a+2b+2c)y-(b+c)z)) + \cos(2\pi((a+2b+2c)x-(b+2c)y+cz)) \right.\\
& \left.+ \cos(2\pi(-(a+b+2c)x+ay+(b+c)z)) - \cos(2\pi((a+b+2c)x+by-(b+c)z)) \right.\\
& \left.- \cos(2\pi(-bx+(a+2b+2c)y-(a+b+c)z)) - \cos(2\pi((a+2b+2c)x-(a+b+2c)y+cz) \right.\\
& \left.- \cos(2\pi(-(b+2c)x-ay+(a+b+c)z)) + \cos(2\pi((b+2c)x+(a+b)y-(a+b+c)z) \right.\\
& \left.- \cos(2\pi((b+2c)x-(a+2b+c)y+(a+b+c)z)) - \cos(2\pi(-(a+2b+2c)x+(a+b)y+cz) \right.\\
& \left.+ \cos(2\pi((a+2b+2c)x-by-cz)) + \cos(2\pi(-(a+b+2c)x+(a+2b+2c)y-(b+c)z) \right\rbrace  .
\end{align*}}
The coefficients $d_\la$ of continuous orthogonality relations \eqref{scorthog} are given in Table \ref{tabStab}. The discrete grid $F_M^s$ is given by
\begin{equation}\label{FMC3}
F^s_M=\set{\dfrac{u_1^s}{M}\o_1^\vee+\dfrac{u_2^s}{M}\o_2^\vee +\dfrac{u_3^s}{M}\o_3^\vee}{u_0^s,u_3^s\in\Z^{\ge0},u_1^s,u_2^s\in\N,u_0^s+2u_1^s+2u_2^s+u_3^s=M }
\end{equation}
and the corresponding finite set of weights has the form
\begin{align}\label{LMC3}\Lambda_M^s&=\set{t_1^s\o_1+t_2^s\o_2+t_3^s\o_3}{t_0^s,t_1^s,t_2^s\in \N,t_0^s\in\Z^{\ge0},t_0^s+t_1^s+2t_2^s+2t_3^s=M}.
\end{align}
The number of points in each of these grids are given as
\begin{align*}
|F_{2k}^s|&=|\Lambda_{2k}^s|=\dfrac{1}{6}k(k-1)(2k-1) \\
|F_{2k+1}^s|&=|\Lambda_{2k+1}^s|=\dfrac{1}{3}k(k+1)(k-1). 
\end{align*}
The grid $F^s_{10}$ of $C_3$ is depicted in Figure \ref{gridC3}.
The coefficients $\ep(x)$ and $h_\la^{\vee }$ of discrete orthogonality relations (\ref{sdortho}) are given in Table \ref{epsilonyC3}. 
{\small\begin{table}
\begin{tabular}{c|c}
$x\in F^s_M$ & $\ep(x)$  \\ \hline
$[u^s_0,u^s_1,u^s_2,u^s_3]$ & $48$  \\
$[0,u^s_1,u^s_2,u^s_3]$ & $24$   \\
$[u^s_0,u^s_1,u^s_2,0]$ & $24$   \\
$[0,u^s_1,u^s_2,0]$ & $12$   \\
\multicolumn{1}{c}{}&\\
$x\in F^l_M$ & $\ep(x)$  \\ \hline
$[u^l_0,u^l_1,u^l_2,u^l_3]$ & $48$  \\
$[u^l_0,0,u^l_1,u^l_2]$ & $24$   \\
$[u^l_0,u^l_1,0,u^l_3]$ & $24$  \\
$[u^l_0,0,0,u^l_3]$ & $8$   \\
\multicolumn{1}{c}{}&\\
\multicolumn{1}{c}{}&\\
\end{tabular}\hspace{12pt}
\begin{tabular}{c|c}
$\la\in \Lambda^s_M$  & $h_\lambda ^{\vee}$ \\ \hline
$[t^s_0,t^s_1,t^s_2,t^s_3]$ & $1$  \\
$[t^s_0,t^s_1,t^s_2,0]$ & $2$   \\
\multicolumn{1}{c}{}&\\
$\la\in \Lambda^l_M$  & $h_\lambda ^{\vee}$ \\ \hline
$[t^l_0,t^l_1,t^l_2,t^l_3]$ & $1$  \\
$[0,t^l_1,t^l_2,t^l_3]$ & $2$  \\
$[t^l_0,0,t^l_2,t^l_3]$ & $2$  \\
$[t^l_0,t^l_1,0,t^l_3]$ & $2$  \\
$[0,0,t^l_2,t^l_3]$ & $4$  \\
$[t^l_0,0,0,t^l_3]$ & $6$  \\
$[0,t^l_1,0,t^l_3]$ & $6$  \\
$[0,0,0,t^l_3]$ & $24$  \\
\end{tabular}\\
\medskip
\caption{The coefficients $\ep(x)$ and $h_\lambda ^{\vee}$ of $C_3$. All variables $u^s_i, t^s_i$ and $u^l_i, t^l_i$, $i=0,1,2,3$, are assumed to be natural numbers.}\label{epsilonyC3}
\end{table}}
Each point $x\in F^s_M$ and each weight $\lambda\in \Lambda^s_M$ are represented by the coordinates $[u^s_0,\,u^s_1,\,u^s_2,\,u^s_3]$ and $[t^s_0,\,t^s_1,\,t^s_2,\,t^s_3]$ from defining equations \eqref{FMC3}, \eqref{LMC3}, respectively. 

\subsubsection{Example of $S^s-$ functions interpolation}
 
\begin{figure}
\begin{center}
\includegraphics[width=5.5in]{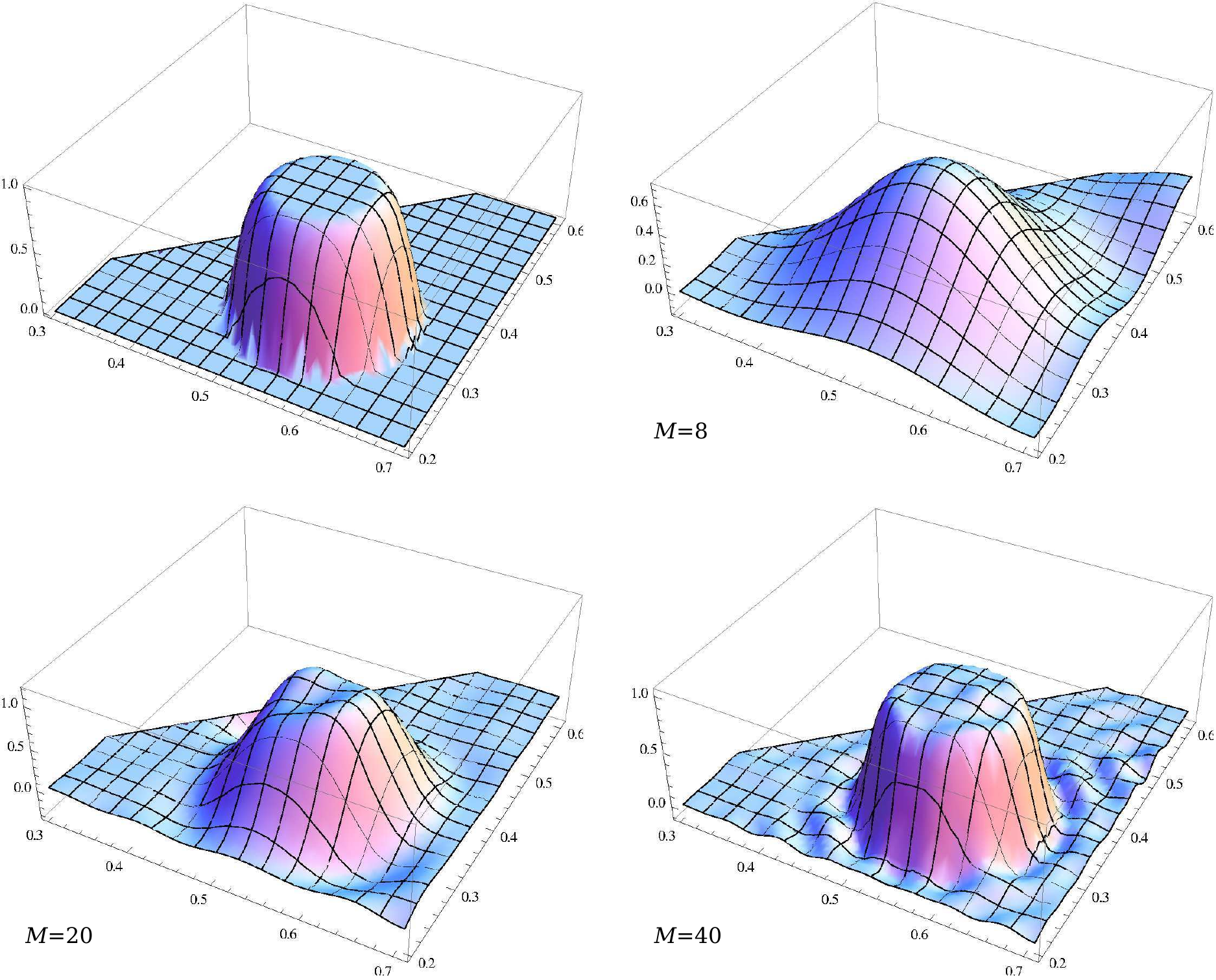}
\caption{The graph cut ($z=\frac{1}{8}$) of the function $f_1$ and the graph cuts of the $f_1-$interpolations $I_M^s$ of $C_3$ for $M=8,\,20$ and $40$.}\label{interpolace2}
\end{center}
\end{figure}

Consider an arbitrary $M\in \N$ and let $f$ be a function sampled on the grid $F_M^s$. An $f-$interpolating function $I_M^s:\R^3\map\C$ is defined as
\begin{equation}\label{iml}
I_M^s(x)=\sum_{\la\in\Lambda_M^s}c_\la^s\varphi_\la^s(x),
\end{equation}
where coefficients $c_\la^s\in \C$ are calculated from \eqref{disc_transform}. As a specific example of a model function, consider the following smooth characteristic function
\begin{equation}\label{interpolfunkce}
f_{\a,\beta,x_0,y_0,z_0}(x,y,z)=\begin{cases} 1&\quad \text{if }r<\a\,,\\ 0&\quad \text{if }r>\beta\,,\\
e \exp\left( \left( \dfrac{r-\a}{\beta-\a}\right)^2-1 \right)^{-1} &\quad \text{otherwise}\,,\\\end{cases}
\end{equation}
where $\a,\beta,x_0,y_0,z_0 \in \R$ and $r=\sqrt{\left(x_0-x \right)^2+ \left(y_0-y \right)^2+\left(z_0-z \right)^2}$.
The parameters in \eqref{interpolfunkce} are set to the values $(\a,\beta)=(\frac{1}{20},\frac{1}{9})$ and $(x_0,y_0,z_0)=(\frac{11}{20},\frac{1}{3},\frac{1}{8})$. The resulting function is denoted by $f_1$ and sampled on the grid $F_M^s$ of $C_3$. 
Fixing the third coordinate $z=\frac{1}{8}$, Figure~\ref{interpolace2} shows graph cuts of $f_1$ and the $f_1-$interpolating functions $I_M^s$ for $M=8,\,20$ and $40$. Integral error estimates
of the form $\int_{F^s} |f_1-I^s_M|^2 \,\mathrm{d}x$ are listed in Table~\ref{odchylky1}.
 
\subsection{$S^l-$functions}\

Choosing $\sigma=\sigma^l$, so--called $S^l-$functions \cite{MMP} are obtained from \eqref{genorb}. Following \cite{HMP}, these functions are here denoted by the symbol $\varphi^l_\la\equiv \psi^{\sigma^l}_\la$. The antiinvariance \eqref{geninv1} with respect to the long reflections $r_\a,\, \a\in \Delta_l$ together with shift invariance \eqref{genshift}
imply zero values of $S^l-$functions on the boundary $H^l$,
$$\varphi^l_{\la}(a')=0,\quad  a'\in H^l.$$
Therefore, the functions $\varphi^l_{\la}$ are considered on the fundamental domain $F^l=F\setminus H^l$ only.
Similarly, the antiinvariance \eqref{geninv2} restricts the weights $\la\in P$ to the set $P^{+l}$.
Thus, we have
\begin{equation*}
\varphi^l_\la(x)=\sum_{w\in W} \sigma^l (w)  \, e^{2 \pi i \sca{ w\la}{x}},\qquad x\in F^l,\, \la \in P^{+l}.
\end{equation*} 

\subsubsection{Continuous orthogonality and $S^l-$transforms}
For any two weights $\la,\la'\in P^{+l}$ the
corresponding $S^l-$functions are orthogonal on $F^l$
\begin{equation}\label{lcorthog}
\int_{F^l}\varphi^l_{\lambda}(x)\overline{\varphi^l_{\lambda'}(x)}\,\mathrm{d}x=
K\,d_{\la}\,\delta_{\lambda\lambda'} \end{equation}
where $d_{\la}$, $K$ are given by \eqref{dla}, \eqref{K}, respectively. The $S^l-$functions determine symmetrized Fourier series expansions,
\begin{equation*}
f(x)=\sum_{\la\in P^{+l}}c^l_{\la}\varphi^l_{\la}(x),\quad {\mathrm{
where}}\
c^l_{\la}=\frac{1}{K d_{\la}}\int_{F^l}f(x)\overline{\varphi^l_{\la}(x)}\,\mathrm{d}x.
\end{equation*}

\subsubsection{Discrete orthogonality and discrete $S^l-$transforms}\ The finite set of points is given by $$F_M^l=\frac{1}{M}P^\vee / Q^\vee \cap F^l$$
and the corresponding finite set of weights as 
$$\Lambda^l_M = P/MQ \cap MF^{l\vee}.$$
Then, for $\la,\la' \in\Lambda^l_M$, the following discrete orthogonality relations hold,
\begin{equation}\label{ldortho}
 \sum_{x\in F^l_M}\ep(x) \varphi^l_\la(x)\overline{\varphi^l_{\la'}(x)}=kM^3 h^{\vee}_\la \delta_{\la\la'}
\end{equation}
where $\ep(x)$, $h^{\vee}_{\la}$ and $k$ are given by \eqref{epx}, \eqref{hla} and \eqref{k}, respectively. The discrete symmetrized $S^l-$function expansion is given by       \begin{equation}\label{disc_transforml}
f(x)=\sum_{\la\in \Lambda_M^l}c^l_{\la}\varphi^l_{\la}(x),\quad {\mathrm{
where}}\
c^l_{\la}=\frac{1}{k M^3 h^{\vee}_\la}\sum_{x\in F^l_M}\ep(x) f(x)\overline{\varphi^l_{\la}(x)}.
\end{equation}

\subsubsection{$S^l-$functions of $B_3$}
For a point with coordinates in $\alpha^\vee-$basis $(x,y,z)$ and a weight with coordinates in $\o-$basis of $(a,b,c)$, the coressponding $S^l-$functions are explicitly evaluated as
{\small\begin{align*}
\varphi^l_{\la}&(x,y,z)
=2\left\lbrace \cos(2\pi(ax+by+cz)) - \cos(2\pi(-ax+(a+b)y+cz))\right.\\  
& \left.- \cos(2\pi((a+b)x-by+(2b+c)z)) + \cos(2\pi(ax+(b+c)y-cz))\right.\\ 
& \left.+ \cos(2\pi(bx-(a+b)y+(2a+2b+c)z)) - \cos(2\pi(-ax+(a+b+c)y-cz))\right.\\
& \left.+ \cos(2\pi(-(a+b)x+ay+(2b+c)z)) - \cos(2\pi((a+b)x+(b+c)y-(2b+c)z)) \right.\\
& \left.- \cos(2\pi((a+b+c)x-(b+c)y+(2b+c)z)) - \cos(2\pi(-bx-ay-(2a+2b+c)z)) \right.\\
& \left.+ \cos(2\pi(bx+(a+b+c)y-(2a+2b+c)z)) + \cos(2\pi((b+c)x-(a+b+c)y+(2a+2b+c)z)) \right.\\
& \left.+ \cos(2\pi(-(a+b)x+(a+2b+c)y-(2b+c)z)) + \cos(2\pi((a+2b+c)x-(b+c)y+cz)) \right.\\
& \left.+ \cos(2\pi(-(a+b+c)x+ay+(2b+c)z)) - \cos(2\pi((a+b+c)x+by-(2b+c)z)) \right.\\
& \left.- \cos(2\pi(-bx+(a+2b+c)y-(2a+2b+c)z)) - \cos(2\pi((a+2b+c)x-(a+b+c)y+cz) \right.\\
& \left.- \cos(2\pi(-(b+c)x-ay+(2a+2b+c)z)) + \cos(2\pi((b+c)x+(a+b)y-(2a+2b+c)z) \right.\\
& \left.- \cos(2\pi((b+c)x-(a+2b+c)y+(2a+2b+c)z)) - \cos(2\pi(-(a+2b+c)x+(a+b)y+cz) \right.\\
& \left.+ \cos(2\pi((a+2b+c)x-by-cz)) + \cos(2\pi(-(a+b+c)x+(a+2b+c)y-(2b+c)z) \right\rbrace .
\end{align*}}
The coefficients $d_\la$ of continuous orthogonality relations \eqref{lcorthog} are given in Table \ref{tabStab}. The discrete grid $F_M^l$ is given by
\begin{equation}\label{FMB3l}
F^l_M=\set{\dfrac{u_1^l}{M}\o_1^\vee+\dfrac{u_2^l}{M}\o_2^\vee +\dfrac{u_3^l}{M}\o_3^\vee}{u_0^l,u_1^l,u_2^l\in\N,u_3^l\in\Z^{\ge0},u_0^l+u_1^l+2u_2^l+2u_3^l=M }
\end{equation}
and the corresponding finite set of weights has the form
\begin{align}\label{LMB3l}\Lambda_M^l&=\set{t_1^l\o_1+t_2^l\o_2+t_3^l\o_3}{t_0^l,t_3^l\in \Z^{\ge0},t_1^l,t_2^l\in\N,t_0^l+2t_1^l+2t_2^l+t_3^l=M}.
\end{align}
The number of points in each of these grids are given as
\begin{align*}
|F_{2k}^l|&=|\Lambda_{2k}^l|=\dfrac{1}{6}k(k-1)(2k-1) \\
|F_{2k+1}^l|&=|\Lambda_{2k+1}^l|=\dfrac{1}{3}k(k+1)(k-1). 
\end{align*}
The grid $F^l_{10}$ of $B_3$ is depicted in Figure \ref{gridB3}.
The coefficients $\ep(x)$ and $h_\la^{\vee }$ of discrete orthogonality relations (\ref{ldortho}) are given in Table \ref{epsilonyB3}. Each point $x\in F^l_M$ and each weight $\lambda\in \Lambda^l_M$ are represented by the coordinates $[u^l_0,\,u^l_1,\,u^l_2,\,u^l_3]$ and $[t^l_0,\,t^l_1,\,t^l_2,\,t^l_3]$ from defining equations \eqref{FMB3l}, \eqref{LMB3l}, respectively.  

\subsubsection{$S^l-$functions of $C_3$}
For a point with coordinates in $\alpha^\vee-$basis $(x,y,z)$ and a weight with coordinates in $\o-$basis of $(a,b,c)$, the coressponding $S^l-$functions are explicitly evaluated as
{\small\begin{align*}
\varphi^l_{\la}&(x,y,z)
=2i\left\lbrace \sin(2\pi(ax+by+cz)) + \sin(2\pi(-ax+(a+b)y+cz))\right.\\  
& \left.+ \sin(2\pi((a+b)x-by+(b+c)z)) - \sin(2\pi(ax+(b+2c)y-cz))\right.\\ 
& \left.+ \sin(2\pi(bx-(a+b)y+(a+b+c)z)) - \sin(2\pi(-ax+(a+b+2c)y-cz))\right.\\
& \left.+ \sin(2\pi(-(a+b)x+ay+(b+c)z)) - \sin(2\pi((a+b)x+(b+2c)y-(b+c)z)) \right.\\
& \left.- \sin(2\pi((a+b+2c)x-(b+2c)y+(b+c)z)) + \sin(2\pi(-bx-ay-(a+b+c)z)) \right.\\
& \left.- \sin(2\pi(bx+(a+b+2c)y-(a+b+c)z)) - \sin(2\pi((b+2c)x-(a+b+2c)y+(a+b+c)z)) \right.\\
& \left.- \sin(2\pi(-(a+b)x+(a+2b+2c)y-(b+c)z)) - \sin(2\pi((a+2b+2c)x-(b+2c)y+cz)) \right.\\
& \left.- \sin(2\pi(-(a+b+2c)x+ay+(b+c)z)) + \sin(2\pi((a+b+2c)x+by-(b+c)z)) \right.\\
& \left.- \sin(2\pi(-bx+(a+2b+2c)y-(a+b+c)z)) - \sin(2\pi((a+2b+2c)x-(a+b+2c)y+cz) \right.\\
& \left.- \sin(2\pi(-(b+2c)x-ay+(a+b+c)z)) + \sin(2\pi((b+2c)x+(a+b)y-(a+b+c)z) \right.\\
& \left.- \sin(2\pi((b+2c)x-(a+2b+c)y+(a+b+c)z)) - \sin(2\pi(-(a+2b+2c)x+(a+b)y+cz) \right.\\
& \left.+ \sin(2\pi((a+2b+2c)x-by-cz)) + \sin(2\pi(-(a+b+2c)x+(a+2b+2c)y-(b+c)z) \right\rbrace  .
\end{align*}}
The coefficients $d_\la$ of continuous orthogonality relations \eqref{lcorthog} are given in Table \ref{tabStab}.
The discrete grid $F_M^l$ is given by
\begin{equation}\label{FMC3l}
F^l_M=\set{\dfrac{u_1^l}{M}\o_1^\vee+\dfrac{u_2^l}{M}\o_2^\vee +\dfrac{u_3^l}{M}\o_3^\vee}{u_0^l,u_3^l\in\N,u_1^l,u_2^l\in\Z^{\ge0},u_0^l+2u_1^l+2u_2^l+u_3^l=M }
\end{equation}
and the corresponding finite set of weights has the form
\begin{align}\label{LMC3l}\Lambda_M^l&=\set{t_1^l\o_1+t_2^l\o_2+t_3^l\o_3}{t_0^l,t_1^l,t_2^l\in \Z^{\ge0},t_3^l\in\N,t_0^l+t_1^l+2t_2^l+2t_3^l=M}.
\end{align}
The number of points in each of these grids are given as
\begin{align*}
|F_{2k}^l|&=|\Lambda_{2k}^l|=\dfrac{1}{6}k(k+1)(2k+1) \\
|F_{2k+1}^l|&=|\Lambda_{2k+1}^l|=\dfrac{1}{3}k(k+1)(k+2). 
\end{align*}
The grid $F^l_{10}$ of $C_3$ is depicted in Figure \ref{gridC3}.
The coefficients $\ep(x)$ and $h_\la^{\vee }$ of discrete orthogonality relations (\ref{ldortho}) are given in Table \ref{epsilonyC3}. Each point $x\in F^l_M$ and each weight $\lambda\in \Lambda^l_M$ are represented by the coordinates $[u^l_0,\,u^l_1,\,u^l_2,\,u^l_3]$ and $[t^l_0,\,t^l_1,\,t^l_2,\,t^l_3]$ from defining equations \eqref{FMC3l}, \eqref{LMC3l}, respectively.  

\subsubsection{Example of $S^l-$ functions interpolation}
 
\begin{figure}
\begin{center}
\includegraphics[width=5.5in]{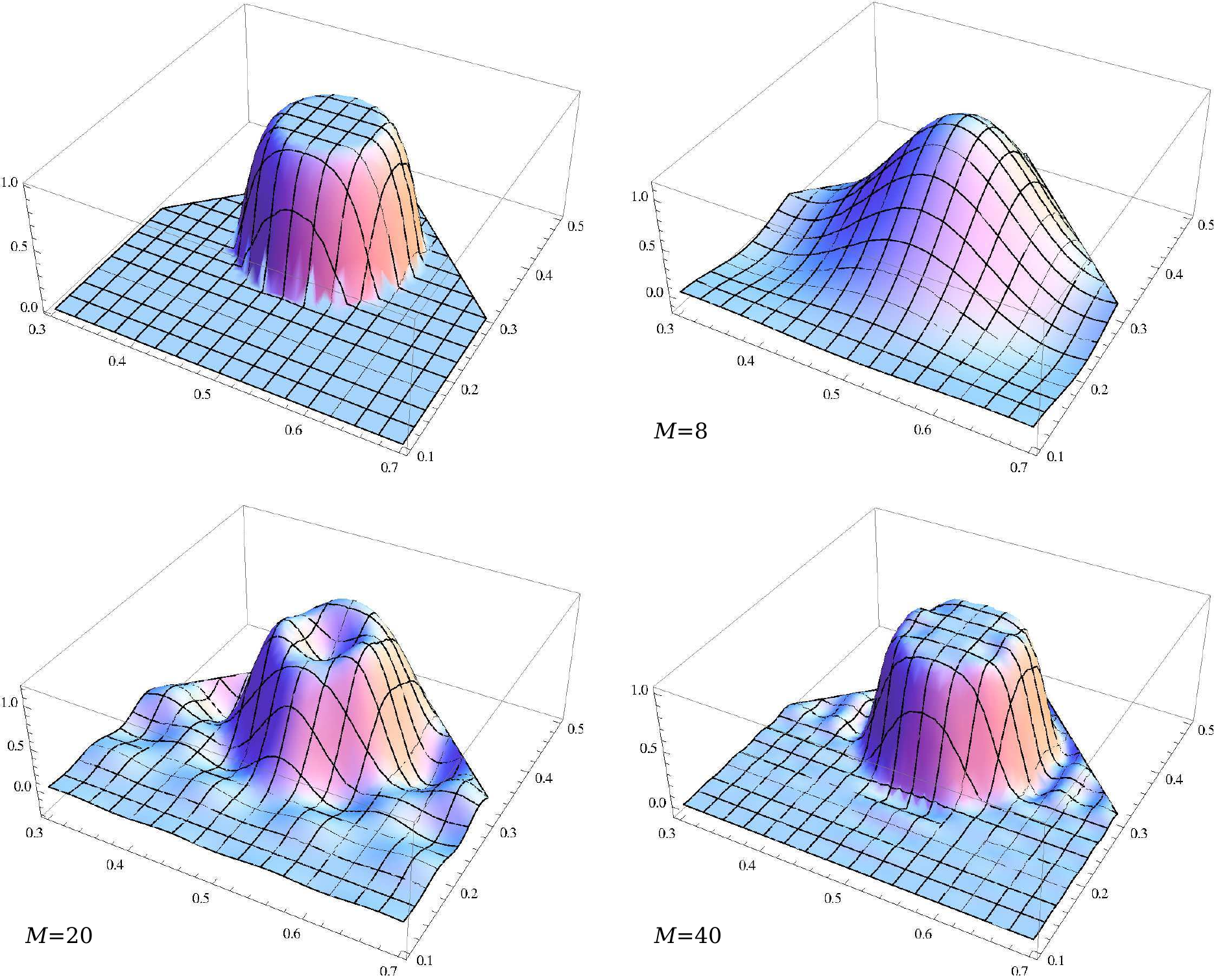}
\caption{The graph cut ($z=\frac{1}{8}$) of the function $f_2$ and the graph cuts of the $f_2-$interpolations $I_M^l$ of $B_3$ for $M=8,\,20$ and $40$.}\label{interpolace}
\end{center}
\end{figure}

Consider an arbitrary $M\in \N$ and let $f$ be a function sampled on the grid $F_M^l$. An $f-$interpolating function $I_M^l:\R^3\map\C$ is defined as
\begin{equation}\label{imll}
I_M^l(x)=\sum_{\la\in\Lambda_M^l}c_\la^l\varphi_\la^l(x),
\end{equation}
where coefficients $c_\la^l\in \C$ are calculated from \eqref{disc_transforml}. The parameters in \eqref{interpolfunkce} are set to the values $(\a,\beta)=(\frac{1}{20},\frac{1}{9})$ and $(x_0,y_0,z_0)=(\frac{1}{2},\frac{1}{3},\frac{1}{8})$ and the resulting function, denoted by $f_2$, is sampled on the grid $F_M^l$ of $B_3$. 
Fixing the third coordinate $z=\frac{1}{8}$, Figure~\ref{interpolace} shows graph cuts of $f_2$ and the $f_2-$interpolating functions $I_M^l$ for $M=8,\,20$ and $40$. Integral error estimates
 of the form
$
\int_{F^l} |f_2-I^l_M|^2 \,\mathrm{d}x
$
are listed in Table~\ref{odchylky1}.

\begin{table}
\begin{tabular}{|c||c|c|}
\hline
$M$ & $\int_{F^s} |f_1-I^s_M|^2 \,\mathrm{d}x$&$ \int_{F^l} |f_2-I^l_M|^2 \,\mathrm{d}x$\\
\hline\hline
8  &$2162,5\times 10^{-6}$& $574,87\times 10^{-6}$\\
16 &$350,62\times 10^{-6}$& $202,74\times 10^{-6}$\\
24 &$77,45\times 10^{-6}$&  $57,16 \times 10^{-6}$\\
32 &$32,14\times 10^{-6}$&  $13,07 \times 10^{-6}$\\
40 &$15,88\times 10^{-6}$&  $12,73 \times 10^{-6}$\\
\hline
\end{tabular}
\medskip
\caption{Integral error estimates of the interpolations $I^s_M$ ans $I^l_M$.}\label{odchylky1}
\end{table}

%%%%%%%%%%%%%%%%%%%%%%%%%%%%%%%%%%%%%%%%%%%%%%%%%%%%%%%%%%%%%%%%
\section{Concluding Remarks}
\begin{itemize}
\item $C-$ and $S-$functions are related by the well known Weyl character formula~\cite{Hum} for irreducible representations of compact semisimple Lie groups. Let $\chi_\lambda$ be the character of an irreducible representation of the group $G$ with the highest root $\la$. Then we have 
\begin{equation*}
\chi_\lambda(x) = \sum_{\mu\in P_+} m_{\mu}^\la \Phi_\mu(x)=\dfrac{\varphi_{\la+\rho}(x)}{\varphi_\rho(x)}\,, 
\end{equation*}
where $\Phi_\la$ and $\varphi_\la$ are the orbit functions related to the Weyl group $W$ of $G$, $\rho $ is the half sum of positive roots in the root system of $W$. The numbers $m_{\mu}^\la$ are the multiplicities of the dominant weights $\mu$ in the weight system of the irreducible representations with the highest weight $\la$. Extensive tables of the multiplicities for simple Lie groups up to rank 12 can be found in~\cite{MP84}.
\item The product of two $S^s-$functions or two $S^l-$functions of $B_3$ or of $C_3$, with the same arguments $x\in\R^3$ but with different weights $\lambda$ and $\lambda'$, decomposes into the sum of $C-$functions:
\begin{equation*}
\varphi^s_\la(x)\cdot \varphi^s_{\la'}(x)=\sum_{w\in W}\sigma^s(w) \Phi_{\la+w\la'}(x),\qquad \varphi^l_\la(x)\cdot \varphi^l_{\la'}(x)=\sum_{w\in W}\sigma^l(w) \Phi_{\la+w\la'}(x).
\end{equation*}
\item The two examples of interpolation show promising behaviour of the $f-$interpolating functions $I^s_M$ and $I^l_M$.
Visual inspection as well as the integral error estimates of the interpolations of the model functions $f_1$ and $f_2$ lead to
the qualitative conclusion that the interpolation error is rather small once the minimal distances of
the lattice grids become smaller than the diameters $\beta$ of the $3D$ jumps of the model functions. The existence of general conditions for the convergence of the functional series $\{I^s_M\}_{M=1}^\infty$, $\{I^l_M\}_{M=1}^\infty$ deserves further study.

\item The $C-$, $S-$, $S^s-$, $S^l-$functions of $C_3$ are directly related to the special functions known as (anti)symmetric trigonometric functions \cite{KP4}; similar symmetrized functions are studied in \cite{BSX}.  These four types of functions are denoted in \cite{KP4} by $\sin^\pm _\la (x)$ and $\cos^\pm _\la (x)$ and are obtained as determinants or permanents of the matrices $(\sin 2\pi \la_i x_j)_{i,j=1}^3$ and $(\cos 2\pi \la_i x_j)_{i,j=1}^3$, respectively. Taking the variables $(x_1,x_2,x_3)\equiv x$ and $(\la_1,\la_2,\la_3) \equiv \la$ in the standard orthonormal basis of $C_3$, the following relations hold
$$\Phi_\la (x)=8 \cos^+ _\la (x),\quad \varphi_\la (x)=-8i \sin^- _\la (x),\quad \varphi^s_\la (x)=8 \cos^- _\la (x),\quad \varphi^l_\la (x)=-8i \sin^+ _\la (x).$$
The discrete Fourier calculus of these functions is, however, performed here on different grids than those in \cite{KP4}.
 
\item The $C-$, $S-$, $S^s-$, $S^l-$functions of $B_3$ are also related to the special functions from \cite{LX2}. They can be obtained by adding one additional symmetry to the functions $\mathsf {TC}_{\mathbf k}({\mathbf t})$, $\mathsf {TS}_{\mathbf k}({\mathbf t})$, which are defined in \cite{LX2} as scalar multiples of $C-$ and $S-$ functions of the algebra $A_3$. Taking the variables $(x_1,x_2,x_3)\equiv x$ and $(\la_1,\la_2,\la_3) \equiv \la$ in the standard orthonormal basis of $B_3$ and homogeneous coordinates ${\mathbf t},\, {\mathbf k}$ of $x$ and $\la$, defined by equation (3.2) in \cite{LX2}, we obtain
$$\Phi_\la (x)=24 (\mathsf {TC}_{\mathbf k} ({4 \mathbf t})+\mathsf {TC}_{-\mathbf k} ({4\mathbf t})),\quad \varphi_\la (x)=-24 (\mathsf {TS}_{\mathbf k} ({4 \mathbf t})-\mathsf {TS}_{-\mathbf k} ({4\mathbf t})),$$ $$ \varphi^s_\la (x)=24 (\mathsf {TC}_{\mathbf k} ({4 \mathbf t})-\mathsf {TC}_{-\mathbf k} ({4\mathbf t})),\quad \varphi^l_\la (x)=-24 (\mathsf {TS}_{\mathbf k} ({4 \mathbf t})+\mathsf {TS}_{-\mathbf k} ({4\mathbf t})).$$
 
\item Gaussian cubature formula, built on Chebyshev polynomials of the second kind (equivalently, the characters of $A_1$), is known to lead to optimal interpolation of polynomial functions of one variable \cite{Riv}. Analogous  result extended to $n$-variable functions with underlying $A_n$-symmetry is known \cite{LX}. Recently it has been further extended to simple Lie groups of all types \cite{MP}. A curious question, which has not yet been answered, is about possibilities to build cubature formulas using the hybrid characters of $B_3$, $C_3$ and other higher rank simple Lie groups.

\item Hybrid characters have other intriguing properties, even if at present we may have no application for them.
It is known that characters of finite dimensional representations of compact simple Lie groups contain finite number of conjugacy classes of elements, represented by discrete points in $F$, with the following property. Character values at such points in $F$ are integers for any representation of the Lie group \cite{MP84}. The simplest example  is the identity element. Its characters are the dimensions of the representations. It was shown in \cite{Sz} that this property carries over to the $C$-functions. Hence it carries over to the hybrid characters. At which point it happens and what are the character values at these points?

\end{itemize}
\section*{Acknowledgments}

We gratefully acknowledge the support of this work by the Natural Sciences and Engineering Research Council of Canada and by the Doppler Institute of the Czech Technical University in Prague. JH is grateful for the hospitality extended to him at the Centre de recherches math\'ematiques, Universit\'e de Montr\'eal. JH gratefully acknowledges support by the Ministry of Education of Czech Republic (project MSM6840770039). JP expresses his gratitude for the hospitality of the Doppler Institute.


\begin{thebibliography}{99}

\bibitem{BSX}
H. Berens, H. Schmid, Y. Xu, \textit{Multivariate Gaussian cubature formulae}, Arch. Math. {\bf 64} (1995) 26--32

\bibitem{HHP}
L. H\'akov\'a, J. Hrivn\'ak, J. Patera, \textit{Six types of $E-$functions of the Lie groups $O(5)$ and $G(2)$} , J. Phys. A: Math. Theor. \textbf{45} (2012) 125201; arXiv:1202.5031

\bibitem{HO}
G. J. Heckman,  E. M. Opdam, \textit{Root systems and hypergeometric functions. I.\/} Compositio Math. {\bf 64} no. 3, (1987), 329--352

\bibitem{HMP}
J. Hrivn\'ak, L. Motlochov\'a, J. Patera, \textit{On discretization of tori of compact
simple Lie groups II},  J. Phys. A: Math. Theor. \textbf{45} (2012) 255201; arXiv:1206.0240

\bibitem{HP}
J. Hrivn\'ak, J. Patera, \textit{On discretization of tori of compact
simple Lie groups,} J. Phys. A: Math. Theor.~{\bf 42} (2009)
385208; arXiv:0905.2395

\bibitem{LX2}
H. Li, Y. Xu, \textit{Discrete Fourier analysis on a dodecahedron and a tetrahedron}, Math. Comput. {\bf 78} (2009) 999--1029; arXiv:0803.0508

\bibitem{LX}
H. Li, Y. Xu, \textit{Discrete Fourier analysis on fundamental domain and simplex of $A_d$ lattice in $d-$variables}, J. Fourier  Anal. Appl. {\bf 16} (2010) 383--433; arXiv:0809.1079


\bibitem{Hum}
J.~E.~Humphreys, \textit{Introduction to Lie Algebras and
Representation Theory,\/} New York, Springer, 1972.

\bibitem{Kane}
R.~Kane, \textit{Reflection Groups and Invariant Theory,\/} New York,
Springer, 2001.

\bibitem{KP2}
A. Klimyk, J. Patera, \textit{Orbit functions,\/}  SIGMA (Symmetry,
Integrability and Geometry: Methods and Applications) {\bf 2}
(2006), 006; arXiv:math-ph/0601037

\bibitem{KP3}
A. Klimyk, J. Patera, \textit{Antisymmetric orbit functions,\/} SIGMA
(Symmetry, Integrability and Geometry: Methods and Applications)
{\bf 3} (2007), 023; arXiv:math-ph/0702040

\bibitem{KP4}
A. Klimyk, J. Patera, \textit{(Anti)symmetric multidimensional trigonometric functions and the corresponding Fourier transforms,\/} J.~Math, Phys. {\bf 48} (2007) 093504; arXiv:0705.4186

\bibitem{KP5}
A. Klimyk, J. Patera, \textit{(Anti)symmetric multidimensional exponential functions and the corresponding Fourier transforms,\/}  J.~Phys.~A: Math. Theor.  {\bf 40} (2007), 10473--10489; arXiv:0705.3572

\bibitem{LPS}
F. Lemire,  J. Patera, M. Szajewska \textit{Decompositions of Hybrid Characters,\/} arXiv:1205.0904 

\bibitem{MMP}
R. V. Moody, L. Motlochov\'a, and J. Patera, \textit{New families of Weyl group orbit functions,\/} arXiv:1202.4415

\bibitem{MP1}
R. V.~Moody, J.~Patera, \textit{Orthogonality within the families of \
$C-$, $S-$, and $E-$functions of any compact semisimple Lie
group,\/} SIGMA (Symmetry, Integrability and Geometry: Methods and
Applications) {\bf 2} (2006) 076, arXiv:math-ph/0611020

\bibitem{MP84}  
R.~V.~Moody, J.~Patera, \textit{Characters of elements of finite order in simple Lie groups,\/} SIAM J. on Algebraic and Discrete Methods {\bf 5} (1984) 359--383

\bibitem{MP2}
R.~V. Moody, J. Patera, \textit{Computation of character
decompositions of class  functions on compact semisimple Lie
groups,\/} Mathematics of Computation {\bf 48} (1987) 799--827

\bibitem{MP}
R. V. Moody, J. Patera, \textit{Cubature formulae for orthogonal polynomials in terms of elements of finite order of compact simple Lie groups,\/} Advances of Applied Math. {
\bf 47} (2011) 509--535; arXiv:1005.2773

\bibitem{PZ1}
J.~Patera, A.~Zaratsyan, \textit{Discrete and continuous cosine
transform generalized to Lie groups SU(3) and G(2),\/} J. Math.
Phys. {\bf 46} (2005) 113506

\bibitem{PZ2}
J.~Patera, A.~Zaratsyan, \textit{Discrete and continuous cosine
transform generalized to the Lie groups $SU(2)\times SU(2)$ and
$O(5)$,\/} J.~Math. Phys. {\bf 46} (2005) 053514

\bibitem{PZ3}
J.~Patera, A.~Zaratsyan, \textit{Discrete and continuous sine
transform generalized to the semisimple Lie groups of rank two,\/}
J. Math. Phys. {\bf 47} (2006) 043512

\bibitem{Riv}
T.~J.~Rivlin, \textit{Chebyshev polynomials. From approximation theory to algebra and number theory,\/} Second edition. Pure and Applied Mathematics (New York). John Wiley \& Sons, Inc., New York, (1990)


\bibitem{Sz}
M. Szajewska, \textit{Four types of special functions of $G_2$ and their discretization,} Integral Transforms Spec. Funct. 23 (2012), no. 6, 455-472
\end{thebibliography}
\end{document}